\documentclass[12pt,preprint]{aastex}
\usepackage{graphicx}

\begin{document}

\title{Identification of Bursting Water Maser Features in Orion~KL}

\author{Tomoya HIROTA\altaffilmark{1,2}, 
Masato TSUBOI\altaffilmark{3}, 
Kenta FUJISAWA\altaffilmark{4}, 
Mareki HONMA\altaffilmark{1,2}, 
Noriyuki KAWAGUCHI\altaffilmark{2,5}, 
Mi Kyoung KIM\altaffilmark{1}, 
Hideyuki KOBAYASHI\altaffilmark{1,2}, 
Hiroshi IMAI\altaffilmark{6}, 
Toshihiro OMODAKA\altaffilmark{6}, 
Katsunori, M. SHIBATA\altaffilmark{1,2}, 
Tomomi SHIMOIKURA\altaffilmark{7}, 
Yoshinori YONEKURA\altaffilmark{8}}
\email{tomoya.hirota@nao.ac.jp}
\altaffiltext{1}{National Astronomical Observatory of Japan, Mitaka, Tokyo 181-8588, Japan}
\altaffiltext{2}{Department of Astronomical Sciences, The Graduate University for Advanced Studies (SOKENDAI), Mitaka, Tokyo 181-8588, Japan}
\altaffiltext{3}{Institute of Space and Astronautical Science, Japan Aerospace Exploration Agency, Sagamihara, Kanagawa 229-8510, Japan}
\altaffiltext{4}{Faculty of Science, Yamaguchi University, Yamaguchi, Yamaguchi 753-8512, Japan}
\altaffiltext{5}{National Astronomical Observatory of Japan, Oshu, Iwate 023-0861, Japan}
\altaffiltext{6}{Graduate School of Science and Engineering, Kagoshima University, Kagoshima, Kagoshima 890-0065, Japan}
\altaffiltext{7}{Department of Astronomy and Earth Sciences, Tokyo Gakugei University, Koganei, Tokyo 184-8501, Japan}
\altaffiltext{8}{Center for Astronomy, Ibaraki University, Mito, Ibaraki 310-8512, Japan}

\begin{abstract}
In February 2011, a burst event of the H$_{2}$O maser in Orion~KL (Kleinmann-Low object) has started after 13-year silence. 
This is the third time to detect such phenomena in Orion~KL, followed by those in 1979-1985 and 1998. 
We have carried out astrometric observations of the bursting H$_{2}$O maser features in Orion~KL with VERA (VLBI Exploration of Radio Astrometry), a Japanese VLBI network dedicated for astrometry. 
The total flux of the bursting feature at the LSR velocity of 7.58~km~s$^{-1}$ reaches 4.4$\times$10$^{4}$~Jy in March 2011. 
The intensity of the bursting feature is three orders of magnitudes larger than that of the same velocity feature in the quiescent phase in 2006. 
Two months later, another new feature appears at the LSR velocity of 6.95~km~s$^{-1}$ in May 2011, separated by 12~mas north of the 7.58~km~s$^{-1}$ feature. 
Thus, the current burst occurs at two spatially different features. 
The bursting masers are elongated along the northwest-southeast direction as reported in the previous burst in 1998. 
We determine the absolute positions of the bursting features for the first time ever with a submilli-arcsecond (mas) accuracy. 
Their positions are coincident with the shocked molecular gas called the Orion Compact Ridge. 
We tentatively detect the absolute proper motions of the bursting features toward southwest direction. 
It is most likely that the outflow from the radio source~I or another young stellar object interacting with the Compact Ridge is a possible origin of the H$_{2}$O maser burst. 
\end{abstract}


\section{Introduction}

In the context of astronomical masers, an enormous outburst of the H$_{2}$O maser at 22~GHz ($\lambda$=1.3~cm) in the Orion~KL (Kleinmann-Low object) region is one of the most enigmatic phenomena. 
The first H$_{2}$O maser burst in Orion~KL was discovered in 1979 and the active phase continued until 1985 with several flare-up events \citep{matveenko1988, garay1989}. 
Based on the VLBI (very long baseline interferometer) observations, the bursting maser feature was revealed to have east-west elongation which was interpreted as a circumstellar disk \citep{matveenko1988}. 
Followed by the active phase for about 6~years, the quiescent phase lasted from 1985 to 1997. 
In this phase, a jet-like maser feature with the northwest-southeast elongation was identified \citep{matveyenko1998}. 
The second burst was detected at the end of 1997 and monitoring observations were carried out with single-dish radio telescopes and VLBI \citep{omodaka1999, shimoikura2005}. 
The structure in the second burst phase was consistent with that in the adjacent quiescent phase \citep{omodaka1999, matveyenko2004, shimoikura2005}. 
The maximum flux density flared up to an order of 10$^{6}$~Jy in both burst phases. 
The line-of-sight velocity with respect to the local standard of rest (LSR) of the bursting maser was $\sim$ 8~km~s$^{-1}$ and was unchanged in both phases, suggesting a common origin. 

Two different scenarios have been proposed for this outburst; one is related to the existence of a circumstellar disk \citep[e.g.][]{matveenko1988,shimoikura2005} and another is the jet/outflow driven by young stellar object (YSO) in the Orion Nebula \citep{garay1989,matveyenko2004}. 
However, no evidence of either circumstellar disk or the jet/outflow system associated with the bursting maser has been identified, and hence, the origin of this maser burst is still unclear. 

The third-time burst of the H$_{2}$O maser in Orion~KL has started since February 2011 \citep{tolmachev2011}. 
The LSR velocity of the bursting maser is again 7.8~km~s$^{-1}$. 
The burst appears to be possibly periodic with an interval of about 13~years (1985, 1998, and 2011), which could support the hypothesis of the common origin. 

In order to explore the mechanism of this maser burst, we have started astrometric observations of the bursting H$_{2}$O maser in Orion~KL with VERA (VLBI Exploration of Radio Astrometry), a Japanese VLBI network developed for astrometry. 
Here we report the result of our absolute position measurement of the bursting maser features with a submilli-arcsecond (mas) accuracy. 
It allows us to measure the proper motions of the bursting features. 
Further detailed studies on the structure and time variation of the bursting maser features will be undertaken. 

\section{Observations}

Observations of the H$_{2}$O maser ($6_{1 6}$-$5_{2 3}$, 22235.080 MHz) in Orion~KL were carried out on 2011 March 09, May 01, and May 17 with VERA. 
The weather conditions were good at all of the four stations \citep[see Figure 1 of][]{petrov2007}, and the typical system noise temperature was 100-200~K. 
The maximum baseline length was 2270~km and the uniform-weighted synthesized beam size (FWMH) was 1.7~mas$\times$0.9~mas with a position angle of 143~degrees on average. 
We employed the dual beam observation mode, in which Orion~KL and an ICRF source J054138.0-054149 \citep{ma1998} were observed simultaneously. 
The data were recorded onto magnetic tapes at a rate of 1024~Mbps, providing a total bandwidth of 256~MHz in which one IF channel and the rest of 15 IF channels with 16~MHz bandwidth each were assigned to Orion~KL and ICRF~J054138.0-054149, respectively. 
For the H$_{2}$O maser lines, a spectral resolution was set to be 15.625~kHz, corresponding to a velocity resolution of 0.21~km~s$^{-1}$. 
A bright continuum source, ICRF~J053056.4+133155, was observed every 80~minutes for bandpass and delay calibration. 
Calibration and imaging were performed using the NRAO Astronomical Image Processing System (AIPS) software package. 

In addition to the present observations, we reanalyzed the archive data of the previous monitoring observations of Orion~KL \citep{hirota2007} conducted on 2006 July 15 in order to compare the results of the new observations. 
Details of the observation and data analysis are described in \citet{hirota2007}.

\section{Results and Discussion}

\subsection{Images and spectra of the bursting maser features}

\begin{figure}
\begin{center}
\includegraphics[width=15cm]{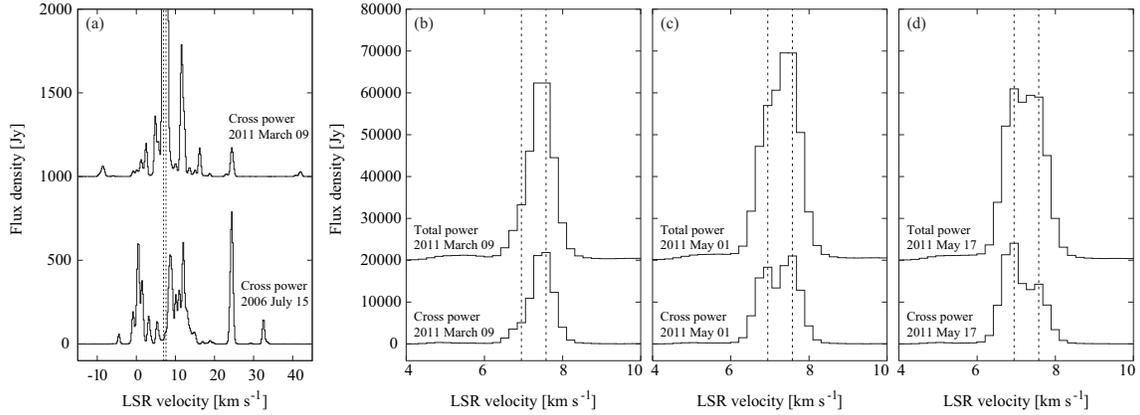}
\caption{Observed spectra of the H$_{2}$O maser in Orion~KL. 
The dashed lines indicate the peak LSR velocity channels at the burst phase, 6.95~km~s$^{-1}$ and 7.58~km~s$^{-1}$. 
(a)  Scalar-averaged cross power spectra at the burst phase on 2011 March 09 (upper) and at the quiescent phase on 2006 July 15 (lower).
The spectra are averaged over all of the time range and all of the baselines. 
(b) Total-power spectrum (i.e. spectrum observed with each single-dish antenna average over the whole of time range and all of the four antennas of VERA) and scalar-averaged cross-power spectrum at the bursting phase on 2011 March 09. 
(c) Same as (b) for the second epoch on 2011 May 01. 
(d) Same as (b) for the third epoch on 2011 May 17. }
\label{fig-sp}
\end{center}
\end{figure}

The H$_{2}$O maser burst is detected at the peak LSR velocity channel of 7.58~km~s$^{-1}$ in the first epoch of observation on 2011 March 09, as shown in Figure \ref{fig-sp}. 
The total flux density is (4.4$\pm$0.3)$\times10^{4}$~Jy. 
The peak velocity is apparently shifted toward lower velocity channel compared with that in the previous burst event, 7.64~km~s$^{-1}$ \citep{shimoikura2005}. 
The correlated flux, $(2.2\pm0.2)\times10^{4}$~Jy, recovers 50\% of the total flux density. 
Due to the lack of short baselines in VERA, extended emission component would be resolved out. 
Two months later from the first epoch, another new velocity component at 6.95~km~s$^{-1}$ appears in 2011 May 01 while the flux density of the 7.58~km~s$^{-1}$ component gradually decreases. 
The total flux density of the 6.95~km~s$^{-1}$ component becomes comparable with that of the 7.58~km~s$^{-1}$ component on 2011 May 17. 

\begin{figure*}
\begin{center}
\includegraphics[width=15cm]{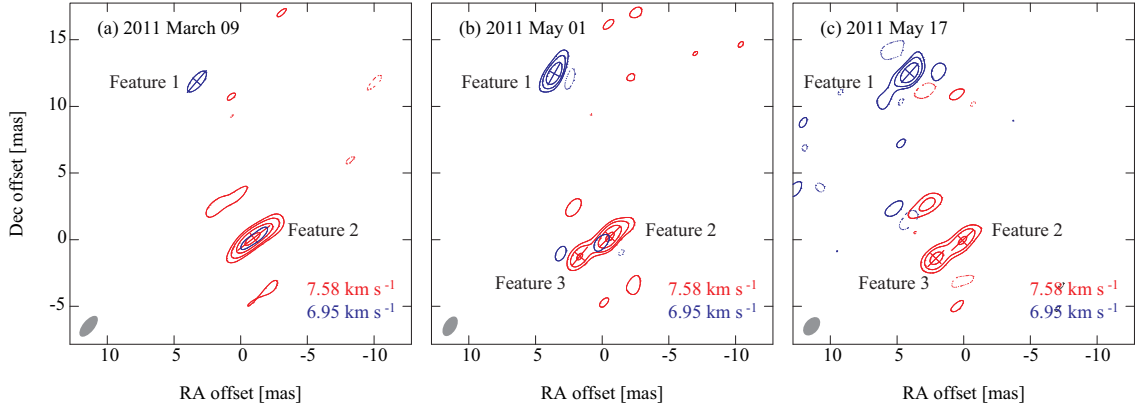}
\caption{Phase-referenced images of the bursting maser features. 
The delay tracking center position (0, 0) is $\alpha$=$05^{\rm{h}}35^{\rm{m}}14^{\rm{s}}.1255$ and $\delta$=$-05^{\circ}22$\arcmin36\arcsec.475 (J2000). 
Results of the Gaussian fitting (major axis, minor axis, and position angle) are indicated with crosses. 
The synthesized beam pattern is indicated by grey ellipse at the bottom left corner in each panel.
(a) The images at the first epoch of the bursting phase on 2011 March 09. 
Red and blue contours represent the 6.95~km~s$^{-1}$ and 7.58~km~s$^{-1}$ features, respectively. 
The contour levels are -1600, 1600, 3200, 6400, and 12800~Jy~beam$^{-1}$. 
(b) Same as (a) but for the second epoch on 2011 May 01. 
 (c) Same as (a) but for the third epoch on 2011 May 17. }
\label{fig-prmap}
\end{center}
\end{figure*}

We have successfully obtained phase-referenced images of the bursting H$_{2}$O maser features. 
The position, peak intensity, and size of each feature are listed in Table \ref{tab-gauss}. 
Note that due to the loss of coherence caused by the atmospheric zenith delay residual \citep[][ and references therein]{hirota2007} in the phase-referencing calibration, the peak flux density is degraded by a factor of 0.5-0.9 compared with the results obtained from the conventional self-calibration method (columns 6 and 7 in Table \ref{tab-gauss}). 
Figure \ref{fig-prmap} shows the phase-referenced images of the bursting maser features at the LSR velocity channels of 6.95~km~s$^{-1}$ and 7.58~km~s$^{-1}$. 
In the first epoch in March 2011, the 7.58~km~s$^{-1}$ feature shows single-peaked structure while it splits into double peaks in May 2011. 
In addition, another spatially distinct feature at the LSR velocity of 6.95~km~s$^{-1}$ appears at 12~mas north as also seen in the spectra (Figure \ref{fig-sp}). 
Therefore, the current maser burst is turned out to occur at two spatially different features. 

The maser features elongate along the northwest-southeast direction. 
Results of the two-dimensional Gaussian fitting on the phase-referenced images are also summarized in Table \ref{tab-gauss}. 
The size of 1~mas corresponds to the linear size of 0.4~AU when adopting the distance to Orion~KL to be 420~pc \citep{hirota2007, kim2008}. 
The structures of the bursting maser features are sometimes resolved along their major axes (i.e. those with the major axes of larger than 2~mas). 
The 7.58~km~s$^{-1}$ feature shows a double-peaked structure with a separation angle of 2.7~mas on 2011 May 01 and 17 as shown in Figures \ref{fig-prmap}(b) and (c). 
The size of the maser feature is comparable to or slightly smaller than that found in the previous burst in 1998 \citep[1.8~mas$\times$0.8~mas-5.3~mas$\times$1.5~mas;][]{shimoikura2005}. 
The brightness temperature reaches 2.3$\times$10$^{13}$~K. 
The velocity width of the total power spectrum of the bursting maser is 0.76$\pm$0.07~km~s$^{-1}$ (FWHM) for the first epoch. 
When we fit the velocity width of each feature at different velocity and position derived from the Gaussian fitting of the self-calibrated images (Table \ref{tab-gauss}), we obtain the average velocity width of 0.61$\pm$0.03~km~s$^{-1}$. 
These velocity widths are broader than that in the previous burst, 0.40-0.48~km~s$^{-1}$ \citep{shimoikura2005}. 

\subsection{Astrometry of the bursting maser features}

In the previous active phases, high-resolution observations were conducted with the VLA and VLBI \citep{matveenko1988, garay1989, greenhill1998, omodaka1999, matveyenko2004, shimoikura2005}. 
With the VLA, the positional accuracy was $\sim$50-100~mas even in the most extended A-configuration \citep[e.g.][]{greenhill1998}. 
Although VLBI observations achieved much higher relative position accuracy of 0.05-0.1~mas, the absolute position information was usually lost by the self-calibration analysis. 

In contrast, we have carried out phase-referencing VLBI astrometry, in which one can measure the position of the target maser source with respect to the extragalactic position reference source. 
For instance, the accuracy of absolute position measurement with VERA is reported to be 0.36~mas and 0.74~mas in right ascension and declination directions, respectively, for the previous H$_{2}$O maser observations in Orion~KL \citep{hirota2007}. 
The main source of the astrometric errors in the VERA observations is the variation of the structures of maser features as well as the atmospheric zenith delay residual due to the troposphere and the positional accuracy of the reference source \citep{ma1998}. 

The observed maser positions are summarized in Table \ref{tab-gauss}. 
This is the first time ever to measure the absolute positions of the bursting maser features in Orion~KL with the submilli-arcsecond accuracy. 
The current position of the bursting maser is shifted by +70~mas and -200~mas in right ascension and declination, respectively, with respect to that of the previous burst in 1983 \citep[e.g.][]{greenhill1998}, probably because of much worse astrometric accuracy of previous observations than that of our present result. 
On the other hand, the position of the bursting maser feature measured in 1998 \citep{omodaka1999} is different by 2\arcsec \ in declination. 
This is due to the lower positional accuracy of an order of $\sim$1\arcsec \ in the fringe-rate mapping method in particular for declination which is typical for the interferometric observations of target sources close to the equator such as Orion~KL. 

\begin{figure}
\begin{center}
\includegraphics[width=15cm]{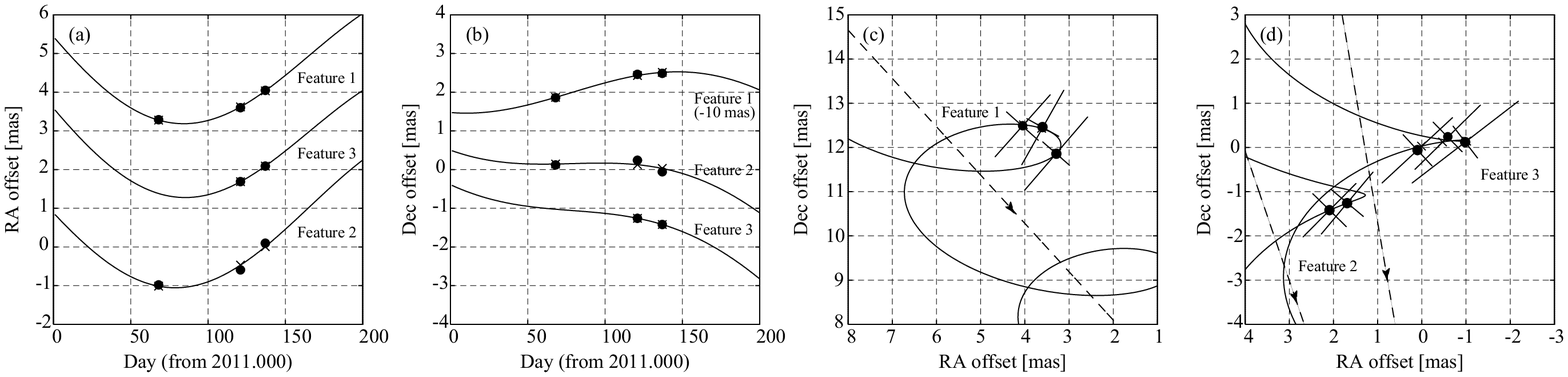}
\caption{Result of position measurements with VERA. 
Filled circles represent the observed positions of the maser features. 
Best-fit models with linear proper motions (as indicated by dashed lines with arrows) and parallax of 2.39~mas \citep{kim2008} are shown by the solid curves. 
Note that the formal errors given in Table \ref{tab-gauss} are as small as 0.03~mas, which cannot be shown in the figure. 
The reference position is the same as that in Figure \ref{fig-prmap}.
(a) Position offset in right ascension as a function of time. 
(b) Same as (a) for declination. 
Declination offset of feature 1 is shifted by -10~mas. 
(c) Movement of the bursting feature 1 on the sky. 
Crosses represent the position and the FWHM size of the maser features listed in Table \ref{tab-gauss}. 
(d) Same as (c) but for the features 2 and 3. 
}
\label{fig-proper}
\end{center}
\end{figure}

We have measured the absolute proper motions of three bursting maser features as shown in Table \ref{tab-proper} and Figure \ref{fig-proper}. 
Taking into account the annual parallax of Orion~KL of 2.39~mas \citep{kim2008}, we obtain the absolute proper motions of $\sim$4-9~mas~yr$^{-1}$ or 8-18~km~s$^{-1}$ toward southwest direction. 
At first glance, they seem inconsistent with the apparent position movement in the phase-referenced images due to the relatively large parallax approximately toward the eastward direction. 
Although the present results contain large uncertainties due to the short monitoring period ($\sim$3~months), the absolute proper motions could be attributed to the outflow in the Orion~KL region as discussed below. 

We tentatively identify the possible counterpart of the bursting maser in the quiescent phase data on 2006 July 15 at the peak LSR velocity channel of 7.37~km~s$^{-1}$, according to the similarity in the line-of-sight velocity, position (within 100~mas from that in March 2011) and the structure (NW-SE elongation). 
The position is also listed in Table \ref{tab-gauss}. 
The peak intensity of the phase-referenced image is only 19.6$\pm$1.4~Jy~beam$^{-1}$ on 2006 July 15 (see Table \ref{tab-gauss}). 
The peak intensity derived from the conventional self-calibrated map is 39.6$\pm$0.5~Jy~beam$^{-1}$. 
The intensity of the bursting feature is three orders of magnitude greater than that of this feature in the quiescent phase. 

Judging from the similarities in peak LSR velocity, position, and structure among the detected features in 2011, 2006, and earlier, they could be attributed to the common origin. 
However, it is unlikely that the bulk of the bursting gas clump would be identical throughout every burst and quiescent phases because the time span of observations ($>$5~yrs) is longer than the typical life time of the H$_{2}$O masers in Orion~KL \citep[$\sim$1~yr; see Figure 1 in ][]{hirota2007}. 
In fact, we can see significant time variation of velocity and spatial structure of the bursting maser features even within the present monitoring observations for only two months. 
If the gas clump of the bursting maser feature would be identical to that of the possible counterpart detected in 2006, one of the features should show an absolute proper motion of (+8~mas~yr$^{-1}$, -17~mas~yr$^{-1}$), which corresponds to the position movement by (+34~mas, -80~mas) in right ascension and declination, respectively, during the 5-years separation. 
We did not detect such a large proper motion in our astrometric data from March to May 2011. 

We also find two other maser features at the same velocity range $\sim$8~km~s$^{-1}$ in the quiescent phase in 2006. 
These positions are shifted by (-1.0\arcsec,-3.1\arcsec) and (+1.7\arcsec, -5.4\arcsec) from the bursting maser features. 
If either of them was identical to the bursting feature, the proper motion would be unrealistically large, 1400-2400~km~s$^{-1}$, according to the position difference during the 5-year separation in our VERA observations. 
Therefore, we conclude that no maser feature detected in 2006 is identical to those in the bursting phase in 2011. 

\subsection{Origin of the bursting maser features}

In this section, we discuss a possible powering source of the bursting maser features. 
One of the most plausible interpretations is related to the outflow from YSO(s) in the Orion~KL region as proposed by \citet{garay1989}. 
It has long been established that the H$_{2}$O masers in Orion~KL are excited by interaction with outflow and ambient dense gas \citep{genzel1981}. 
The most possible powering source is the radio source~I which drives northeast-southwest low-velocity outflow \citep{plambeck2009}. 
The bursting maser features are located at 8\arcsec \ or 3400~AU southwest from source~I, and is coincident with the interacting region between the outflow and a dense ambient gas, which is called the Orion Compact Ridge \citep[e.g.][]{liu2002}. 
Although the elongation of the bursting maser features (PA=130-150~degrees) are perpendicular to the direction to source~I, the maser features could trace the shocked layer between the low-velocity outflow and the Compact Ridge. 
The H$_{2}$O maser burst may occur at the different part of the shocked layer when the episodic or possibly 13~year-periodic outflow from source~I interacts with the Compact Ridge. 
The magnitude of the proper motions of the bursting maser features are 16-20~km~s$^{-1}$ pointing toward the west to southwest direction when we subtract the absolute proper motion of source~I (6.3~mas~yr$^{-1}$, -4.2~mas~yr$^{-1}$) recently reported by \citet{goddi2011}. 
Interestingly, they are roughly consistent with the low-velocity outflow from source~I \citep{genzel1981} although this should be confirmed by the further VLBI monitoring observations. 

Nevertheless, it is still unclear why only the 8~km~s$^{-1}$ features show such anomalous outburst. 
It should be noted that the current burst event cannot be explained by the local phenomenon around each feature with the scale of a few milliarcseconds \citep[e.g.][]{matveenko1988, shimoikura2005} because the maser burst is occurring at two distinct features separated by 12~mas and at different velocities. 
One of the possibilities is that powerful outflows from source~I may interact with the dense core associated with a pre-existing YSO in the Compact Ridge region as suggested by \citet{garay1989}. 
What makes the bursting features always at about 8~km~s$^{-1}$ would be that it is the rest velocity of the Compact Ridge \citep[e.g.][]{liu2002}. 
The low-velocity outflow hitting at any fortuitous place in the Compact Ridge may amplify the maser emission most efficiently at the rest velocity, leading to such features. 
According to the SIMBAD database, there are number of closer infrared, radio and X-ray sources around the bursting maser features than source~I. 
The nearest YSO candidate is the radio continuum source~R, which is 0.5\arcsec \ southwest from the bursting maser features. 
A millimeter and infrared counterpart, HC438, is also associated with the same source \citep{eisner2008}. 
Submillimeter continuum emission peaks are detected within their beam widths \citep{tang2010, zapata2011}. 
Although the mass of the submillimeter core is estimated to be $\sim$4~M$\odot$ \citep{tang2010}, the nature of this source, whether it is a self-luminous YSO or not, is still unclear. 
Thus, we cannot definitely identify one of these sources as a powering source of the bursting maser features only based on its proximity at this moment. 

\section{Summary}
We have successfully measured the absolute positions of the bursting H$_{2}$O maser features in Orion~KL with the VLBI astrometry technique using VERA. 
This is the first time ever to identify the bursting maser features with better than the milli-arcsecond accuracy. 
The burst event is found to be occurring at two spatially different maser features. 
These features are located in the interacting region of the outflow from the radio source~I and the Orion Compact Ridge. 
Although the spatial and velocity structure of the bursting masers are variable with time, the position and structure of the bursting masers are similar to those in the previous burst phases, implying a common origin. 
We suggest that the outflow from source~I or another YSO in the Compact Ridge region could be a possible powering source of the bursting maser features. 

If the current burst phase persists for one year as in the case of previous bursts, we will be able to determine the proper motions of the bursting maser features more accurately through our VLBI astrometry with VERA. 
They will be crucial to reveal dynamical structure of the masing region. 
For instance, we will provide novel information on the VLBI images and proper motions to construct three-dimensional dynamical model of the outflow. 
If the bursting masers are associated with a circumstellar disk, we will be able to derive the mass of the central protostar by directly measuring the rotating motion. 
In addition, high spatial resolution observations with ALMA and EVLA of the thermal molecular lines and continuum emissions from centimeter to submillimeter wavelengths are still indispensable to identify the powering source of the bursting maser along with the velocity structure of possible outflow from it. 
All of these results are essential to shed light on the emission mechanism and origin of the burst of the maser features in Orion~KL, as well as the driving source(s) of complex outflows in this region. 

\acknowledgements
We thank the staff of the VERA stations to support our project, Shuji Deguchi, Satoshi Yamamoto, and anonymous referees for useful comments. 
This work is supported in part by The Graduate University for Advanced Studies (Sokendai), and has made use of the SIMBAD database, operated at CDS, Strasbourg, France. 

{}

\begin{deluxetable}{lccrrrrrrr}
\rotate
\tabletypesize{\scriptsize}
\tablewidth{0pt}
\tablecaption{Properties of the bursting maser features
 \label{tab-gauss}}
\tablehead{
\colhead{}      & \colhead{Feature} & \colhead{$v_{LSR}$}     & \colhead{$\Delta \alpha \cos \delta$} & \colhead{$\Delta \delta$} & \colhead{$I_{\mbox{pr}}$\tablenotemark{b}} & \colhead{$I_{\mbox{sc}}$\tablenotemark{c}} & \colhead{$\theta_{maj}$} & \colhead{$\theta_{min}$} & \colhead{PA} 
 \\
\colhead{Epoch} & \colhead{ID\tablenotemark{a}} & \colhead{(km s$^{-1}$)} & \colhead{(mas)}                       & \colhead{(mas)}           & \colhead{(Jy~beam$^{-1}$)}                 & \colhead{(Jy~beam$^{-1}$)}                 & \colhead{(mas)}          & \colhead{(mas)}          & \colhead{(degree)} }
\startdata
July 15 2006   & \nodata & 7.37 & -34.20$\pm$0.05 & 79.56$\pm$0.05 & (1.96$\pm$0.14)$\times$10$^{1}$ & (3.96$\pm$0.05)$\times$10$^{1}$   & 2.14$\pm$0.15  & 0.76$\pm$0.06 &  139.0$\pm$2.4 \\
March 09 2011   & 1 & 6.95 &   3.29$\pm$0.02 & 11.86$\pm$0.02 & (0.29$\pm$0.01)$\times$10$^{4}$ & (0.378$\pm$0.002)$\times$10$^{4}$ & 2.18$\pm$0.07  & 0.80$\pm$0.03 &  139.0$\pm$1.1 \\
               & 2 & 7.58 &  -0.98$\pm$0.03 &  0.12$\pm$0.02 & (1.52$\pm$0.04)$\times$10$^{4}$ & (1.958$\pm$0.002)$\times$10$^{4}$ & 3.03$\pm$0.08  & 0.90$\pm$0.02 &  127.5$\pm$0.7 \\
May 01 2011   & 1 & 6.95 &   3.60$\pm$0.01 & 12.46$\pm$0.02 & (1.28$\pm$0.03)$\times$10$^{4}$ & (1.567$\pm$0.009)$\times$10$^{4}$ & 1.93$\pm$0.04  & 0.82$\pm$0.02 &  150.7$\pm$1.0 \\
               & 2 & 7.58 &  -0.59$\pm$0.02 &  0.24$\pm$0.02 & (1.45$\pm$0.04)$\times$10$^{4}$ & (1.680$\pm$0.004)$\times$10$^{4}$ & 2.07$\pm$0.05  & 0.93$\pm$0.02 &  134.5$\pm$1.1 \\
               & 3 & 7.58 &   1.69$\pm$0.03 & -1.26$\pm$0.03 & (0.68$\pm$0.04)$\times$10$^{4}$ & (0.819$\pm$0.004)$\times$10$^{4}$ & 1.81$\pm$0.10  & 0.93$\pm$0.05 &  140.0$\pm$3.0 \\
May 17 2011   & 1 & 6.95 &   4.05$\pm$0.02 & 12.49$\pm$0.02 & (1.25$\pm$0.04)$\times$10$^{4}$ & (1.889$\pm$0.016)$\times$10$^{4}$ & 1.87$\pm$0.06  & 0.96$\pm$0.03 &  138.4$\pm$2.0 \\
               & 2 & 7.58 &   0.10$\pm$0.03 & -0.06$\pm$0.03 & (0.71$\pm$0.03)$\times$10$^{4}$ & (1.033$\pm$0.004)$\times$10$^{4}$ & 2.18$\pm$0.10  & 0.96$\pm$0.04 &  132.9$\pm$1.9 \\
               & 3 & 7.58 &   2.09$\pm$0.02 & -1.42$\pm$0.02 & (0.84$\pm$0.03)$\times$10$^{4}$ & (1.098$\pm$0.004)$\times$10$^{4}$ & 1.70$\pm$0.06  & 1.06$\pm$0.04 &  135.6$\pm$3.2 \\
\enddata
\tablenotetext{a}{Feature IDs are shown in Figure \ref{fig-prmap}.}
\tablenotetext{b}{Peak intensity derived from the phase-referencing method. }
\tablenotetext{c}{Peak intensity derived from the self-calibration method. }
\tablecomments{Reference position is $\alpha$=$05^{\rm{h}}35^{\rm{m}}14^{\rm{s}}.1255$ and $\delta$=$-05^{\circ}22$\arcmin36\arcsec.475 (J2000). }
\end{deluxetable}

\begin{deluxetable}{ccrr}
\tabletypesize{\scriptsize}
\tablewidth{0pt}
\tablecaption{Results of the proper motion measurements with VERA
 \label{tab-proper}}
\tablehead{
\colhead{Feature} & \colhead{$v_{LSR}$}     & \colhead{$\mu_{\alpha} \cos \delta$} & \colhead{$\mu_{\delta}$}  \\
\colhead{ID\tablenotemark{a}} & \colhead{(km s$^{-1}$)} & \colhead{(mas~yr$^{-1}$)}  & \colhead{(mas~yr$^{-1}$)} }
\startdata
1 & 6.95 &  -2.6$\pm$0.8 & -2.8$\pm$0.8 \\
2 & 7.58 &  -1.2$\pm$0.8 & -6.8$\pm$0.8 \\
3 & 7.58 &  -2.8$\pm$3.6 & -8.3$\pm$3.6 \\
\enddata
\tablenotetext{a}{Feature IDs are shown in Figure \ref{fig-prmap}.}
\end{deluxetable}

\end{document}